\documentclass[twocolumn,showpacs,preprintnumbers,amsmath,amssymb,prb]{revtex4}

\usepackage{graphicx}% Include figure files
\usepackage{dcolumn}% Align table columns on decimal point
\usepackage{bm}% bold math
\usepackage{color}

\begin{document}

%\preprint{Draft-12: Please don't circulate.}%{APS/123-QED}

\title{Theoretical study of isolated dangling bonds, dangling bond wires and dangling bond clusters on H:Si(100)-(2$\times$1) surface}% Force line breaks with \\

\author{Hassan Raza}
 \email{hraza@purdue.edu}
  %\homepage{http://min.ecn.purdue.edu/~hraza}
   %\altaffiliation{School of Electrical and Computer Engineering, Purdue University, W. Lafayette, IN 47907.}
 \address{NSF Network for Computational Nanotechnology and School of Electrical and Computer Engineering, Purdue University, West Lafayette, Indiana 47907 USA}%

%\date{\today}% It is always \today, today, but any date may be explicitly specified

\begin{abstract}
We theoretically study the electronic band structure of isolated unpaired and paired dangling bonds (DB), DB wires and DB clusters on H:Si(100)-(2$\times$1) surface using Extended H\"uckel Theory (EHT) and report their effect on the Si band gap. An isolated unpaired DB introduces a near-midgap state, whereas a paired DB leads to $\pi$ and $\pi^*$ states, similar to those introduced by an unpassivated asymmetric dimer (AD) Si(100)-(2$\times$1) surface. Such induced states have very small dispersion due to their isolation from the other states, which reside in conduction and valence band. On the other hand, the surface state induced due to an unpaired DB wire in the direction along the dimer row (referred to as $[\overline{1}10]$), has large dispersion due to the strong coupling between the adjacent DBs, being 3.84$\AA$ apart. However, in the direction perpendicular to the dimer row (referred to as [110]), due to the reduced coupling between the DBs being 7.68$\AA$ apart, the dispersion in the surface state is similar to that of an isolated unpaired DB. Apart from this, a paired DB wire in $[\overline{1}10]$ direction introduces $\pi$ and $\pi^*$ states similar to those of an AD surface and a paired DB wire in [110] direction exhibits surface states similar to those of an isolated paired DB, as expected. Besides this, we report the electronic structure of different DB clusters, which exhibit states inside the band gap that can be interpreted as superpositions of states due to unpaired and paired DBs.% leading to finite life time and hence dispersion due to uncertainty principle.} 
%Similarly, an isolated paired DB introduces states within band gap, which are less broadened in energy as compared to unpassivated asymmetric dimer (AD) Si(100)-(2$\times$1). 
\end{abstract}
%73.20.-r 	Electron states at surfaces and interfaces
%73.20.At 	Surface states, band structure, electron density of states
%74.78.Na 	Mesoscopic and nanoscale systems
%79.60.Jv 	Interfaces; heterostructures; nanostructures
%72.10.-d 	Theory of electronic transport; scattering mechanisms
%31.15.Ar Ab initio calculations
%81.07.Nb Molecular Nanostructures
%81.07.Lk Nanocontacts
%85.65.+h Molecular electronic devices
%72.10.Bg General formulation of transport theory
%72.20.Dp General theory, scattering mechanisms of conductivity
%73.23.-b Electronic transport in mesoscopic systems
%73.40.Sx Metal-semiconductor-metal structures
%73.63.-b Electronic transport in mesoscopic or nanoscale materials and structures

\maketitle
\section{Introduction} Due to their importance in Si technology, states induced by the dangling bonds (DBs) on Si(100) surface, known as $P_b$ centers, have been a topic of study for decades\cite{Bardeen47,Wagner72,Uhrberg81,Feenstra86,Hamers86,Hamers88,Wolkow88,Wolkow95,Neergaard95,Hamers96,Hitosugi98,Watanabe96,Muller99,Zhang06,Raza06}. These are important due to their role in introducing states inside the band gap\cite{Bardeen47, Hamers86, Raza06, Watanabe96, Hitosugi98, Muller99}, Fermi level pinning\cite{Hamers93}, charging issue - consider variation of threshold voltage due to charged DBs, reliability issues - consider negative bias temperature instability\cite{Alam04}, enhancement of 1/f noise due to trap states resulting in poor DC characteristics\cite{Butt06}. With advances in nanoscale science, detailed understanding of these surface states is needed to exploit their role in possible technological applications, \textit{e.g.} use of DB as a template for the growth of molecular nanostructures\cite{Wolkow00}. This realization has been the motivation behind research like desorption of H from Si surface forming a DB\cite{Shen95,Stokbro98,Liu06}, electronic and electrostatic effect of DB on transport through a styrene wire\cite{Raza06, Wolkow05}, DB dynamics on Si surface\cite{McEllistrem98}, Jahn-Teller distortion in DB wires at low temperature\cite{Hitosugi99}, magnetism in DB structures\cite{Daniel72,Xiao04,Artacho89}, passivation of DBs with H and D\cite{Hersam02}, etc. 

\begin{figure}
\vspace{3.2in}
%\centering\hskip -2.75in
\hskip -4in\includegraphics{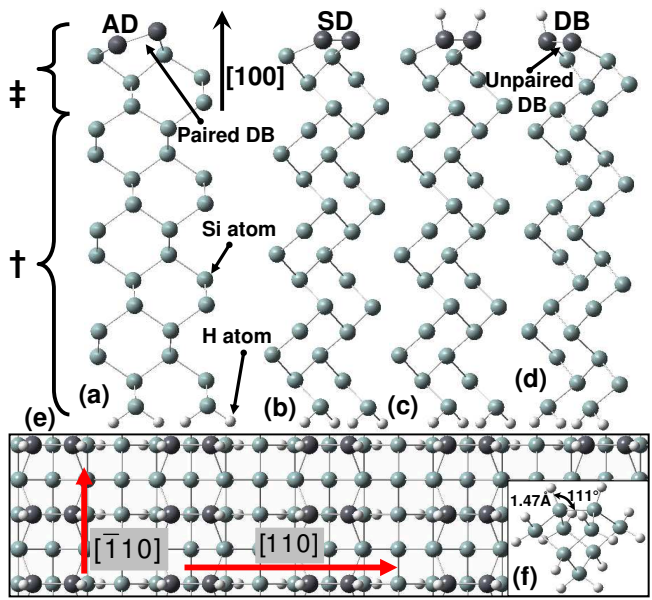}
\caption{(color online) Ball and stick model of sixteen layer Si(100)-(2$\times$1) unit cell used to construct different surfaces. Top four layers, represented by $\ddag$, are relaxed due to surface reconstruction. Bottom twelve, represented by $\dag$, are bulk layers. (a) Unpassivated asymmetric dimer (AD) unit cell referred to as paired dangling bond (DB) configuration. (b) Unpassivated symmetric dimer (SD) unit cell. (c) Hydrogenated SD unit cell. (d) Hydrogenated unit cell with one DB referred to as unpaired DB. Arrow shows [100] direction, perpendicular to the surface. (e) H:Si(100)-(2$\times$1) surface with direction along and perpendicular to dimer row labeled as $[\overline{1}10]$ and  [110] respectively. In part(a-e), back surface is hydrogenated to eliminate any DB induced states. (f) A $Si_9H_{14}$ cluster with H-Si bond length of 1.47$\AA$ and H-Si dimer angle of $111^\circ$.}%Atomic structure for (b) and (c) is taken from Ref. \cite{Ramstad95} and for (a) and (d), SD surface atomic structure is used with added H atom(s). These unit cells are used in constructing the bigger unit cells, as described in Table I and few of them shown in Fig. 2, for doing isolated DB, DB wire and cluster calculations.
\end{figure}

Watanabe \textit{et al}\cite{Watanabe96} predicted that an unpaired DB (Si dimer with one hydrogenated atom) wire on H:Si(100)-(2$\times$1) surface in the direction along the dimer row would introduce states inside the band gap. This prediction was later experimentally observed, however the mechanism is still not clear\cite{Hitosugi98}. Recently, Cakmak \textit{et al}\cite{Cakmak03} theoretically reproduced their results and found similar trend. All these calculations were based on Density Functional Theory (DFT) where band gap is underestimated \cite{SrivastavaBook}. It is thus hard to quantitatively benchmark the results. In this paper, we use Extended H\"uckel Theory (EHT) for electronic band structure calculations using transferable parameters developed by Cerda \textit{et al}\cite{Cerda00}. Notably, EHT gives correct Si bulk band gap and other features as reported by Kienle \textit{et al}\cite{Diego06}. It has been applied to Si surfaces obtaining satisfactory results\cite{Diego06,Liang05}. The state of the art in electronic structure calculations for Si is GW approximation\cite{Hybersten85,Rohlfing93}. However, computational complexity prohibits its use in transport calculations for large systems like the ones discussed in this paper where the Si atoms per unit cell may approach 384. EHT provides computationally accessible, yet accurate approach. Apart from this, to the best of our knowledge there has not yet been a systematic study of isolated DBs, DB wires and DB clusters on H:Si(100)-(2$\times$1) surface using a method that quantitatively addresses the surface states. 

In a previous study\cite{Raza06}, we report that a DB electronically affects Si atoms up to approximately 10$\AA$ away from it, by introducing a near-midgap state in the local density of states of the neighboring Si atoms. Therefore, to simulate DBs in isolated, wired and clustered configurations, we use a large enough unit cell in the directions parallel to and perpendicular to the dimer row, conveniently referred to as $[\overline{1}10]$ and [110] directions respectively \cite{SrivastavaBook} (as shown in Fig. 1(e)). Few large unit cells are shown in Fig. 2 to demonstrate the approach. This approach has been used previously in Ref. \cite{Raza06,Cakmak03,Watanabe96}. Table I summarizes different surfaces being investigated. We use electronic band structure principles in $[\overline{1}10]$ and [110] directions. However in [100] direction, \textit{i.e.} direction normal to surface, we simulate a semi-infinite structure by using sixteen layers of Si atoms as shown in Fig. 1(a-d). The important aspect of this approximation is the use of a large enough unit cell in [100] direction, which minimizes the quantum confinement effects and the correct Si band gap for H:Si(100)-(2$\times$1) is obtained (as shown in Fig. 3(a)). Fig. 3(a) also shows the $\pi$ and $\pi^*$ states for unpassivated asymmetric dimer (AD) and unpassivated SD surface, referred to as Si(100)-(2$\times$1)-AD surface and Si(100)-(2$\times$1)-SD surface respectively. Fig. 3(b) shows transmission through the above-mentioned three surfaces. H:Si(100)-(2$\times$1) surface has a clean band gap, Si(100)-(2$\times$1)-AD surface has 0.62eV of band gap and Si(100)-(2$\times$1)-SD surface has no band gap. %Such quantum confinement would affect the surface states\cite{Watanabe96,Diego06,Liang05} thus obscuring the quantitative information. 

We present calculations of isolated unpaired DB (Si dimer with one atom hydrogenated) and paired DB (unpassivated Si dimer) in Fig. 4(a) and (b) respectively. An unpaired DB introduces a near-midgap state as we previously report\cite{Raza06}, whereas a paired DB introduces a $\pi$ state close to the valence band maximum ($E_v$) and a $\pi^*$ state close to the conduction band minimum ($E_c$). These states due to a paired DB are similar to the $\pi$ and $\pi^*$ states introduced by Si(100)-(2$\times$1)-AD surface as shown in Fig. 3(a), however dispersion is smaller than that of AD surface states. We provide details of bandwidths of different states in Section IV. Furthermore, we model DB wires consisting of unpaired and paired DBs in $[\overline{1}10]$ and [110] directions as shown in Fig. 4(a) and (b) respectively. For unpaired DB wires, in $[\overline{1}10]$ and [110] direction, a near-midgap state similar to that of an isolated unpaired DB is introduced. However, the dispersion in the surface state due to an unpaired DB wire in $[\overline{1}10]$ direction is larger than that of an isolated unpaired DB, whereas the dispersion in surface states due to an unpaired DB wire in [110] direction is only slightly larger. Similarly, for wires consisting of paired DBs in $[\overline{1}10]$ and [110] direction, $\pi$ and $\pi^*$ states are introduced close to $E_v$ and $E_c$ respectively, which are similar to those introduced by an isolated paired DB. The dispersion in surface states of these wires  follow similar pattern as that of unpaired DB wires in $[\overline{1}10]$ and [110] directions respectively. Finally, we report electronic band structure calculations for isolated clusters of DBs as shown in Fig. 5, reporting that electronic structure of clusters can be interpreted as combinations of those of unpaired and paired DBs, but sufficiently dispersed due to interaction within the DBs.

% are much broadened whereas their location stays same as that of isolated paired DB, \textit{i.e.} $\pi$ state close to $E_v$ and $\pi^*$ state close to $E_c$. In [110] direction, dispersion is similar to that of a paired DB as expected. Finally, we report electronic band structure calculations for isolated clusters of DBs as shown in Fig. 5, reporting that electronic structure of clusters can be interpreted as combinations of those of unpaired and paired DBs, although broadened due to interaction with each other.

% due to finite life time leading to dispersion due to uncertainty principle}. %Prior to our study, unpaired DB wires in $[\overline{1}10]$ and [100] directions have been origionally investigated by Watanabe \texit{et al}\cite{Watanabe96} using DFT. For these structures, we obtain similar trends quantitatively and extend on other configurations as shown in Table I. Furthermore, isolated unpaired DB has been studied using EHT as in Ref.\cite{Raza06}. 

This paper is organized in four sections. First we talk about atomic structure and the assumptions made. Then we briefly discuss the theoretical approach. Finally, results and conclusions are presented. 

\begin{table}
\caption{\label{tab:table1}Details of the unit cells used to simulate different surfaces. Sixteen layer atomic structures shown in Fig. 1 are used to construct bigger unit cells. Thus, a (4$\times$6) unit cell has dimensions equal to 15.36$\AA$ and 23.04$\AA$ in [110] (perpendicular to dimer row) and $[\overline{1}10]$ (along dimer row) directions respectively. The total number of Si atoms (excluding H atoms on top and back surface) per unit cell are given.}
\begin{ruledtabular}
\begin{tabular}{llll}
Unit cell surface & Unit cell\footnotemark[1] & Atoms\footnotemark[2]\\
\hline
H:Si(100)-(2$\times$1) & 2$\times$1 & 32 \\
Si(100)-(2$\times$1)-AD & 2$\times$1 & 32\\
Si(100)-(2$\times$1)-SD & 2$\times$1 & 32\\
Unpaired DB & 4$\times$4 & 256\\
Unpaired DB wire\footnotemark[3] & 4$\times$1 & 64\\
Unpaired DB wire\footnotemark[4] & 2$\times$4 & 128\\
Paired DB & 4$\times$4 & 256\\
Paired DB wire\footnotemark[3] & 6$\times$1 & 96\\
Paired DB wire\footnotemark[4] & 2$\times$4 & 128\\
Unpaired and paired DBs\footnotemark[3] & 4$\times$5 & 320\\
Two Paired DBs\footnotemark[4] & 4$\times$5 & 320\\
Paired, Unpaired and paired DBs \footnotemark[3] & 4$\times$6 & 384\\
Unpaired, paired and unpaired DBs\footnotemark[3] & 4$\times$6 & 384\\
\end{tabular}
\end{ruledtabular}
\footnotetext[1]{Multiples of 3.84$\AA$ - lattice constant of bulk unit cell.}
\footnotetext[2]{Number of Si atoms only per unit cell.}
\footnotetext[3]{in $[\overline{1}10]$ direction.}
\footnotetext[4]{in [110] direction.}
\end{table}

\begin{figure*}
\vspace{4in}
%\centering\hskip -3.5in
\hskip -5.5in\includegraphics{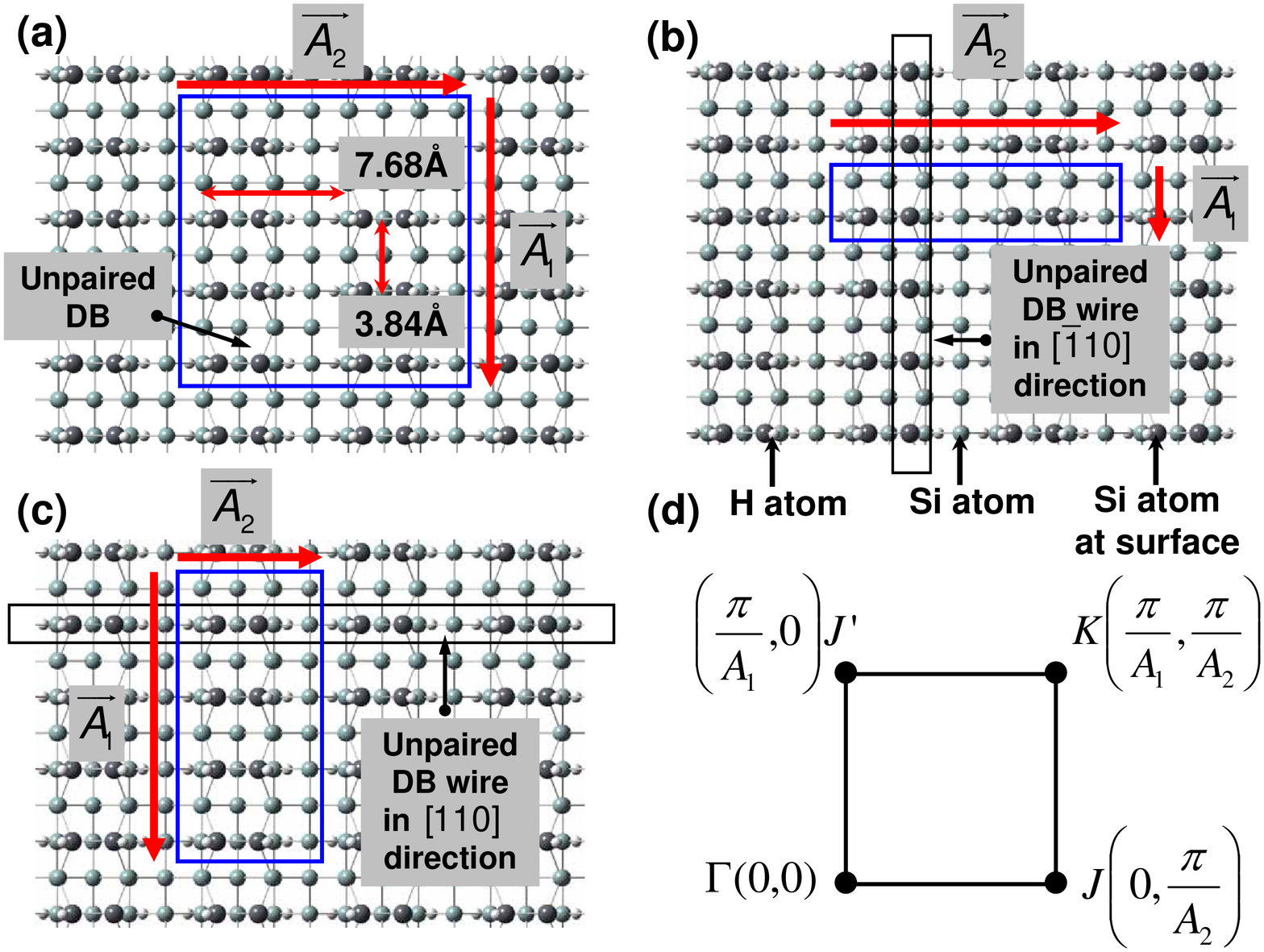}
\caption{(color online) Ball and stick model for the unit cells of different configurations shown to demonstrate the approach used to calculate the electronic structure. The dimensions of unit cells (in $[\overline{1}10]$ and [110] direction) for isolated DB, DB wire and DB cluster are adopted in such a way that the DB in one unit cell does not interfere with the neighboring unit cells in any inadvertent way. Note that $A_1$ and $A_2$ are the unit vectors in $[\overline{1}10]$ and [110] directions respectively. (a) Isolated unpaired DB on H:Si(100)-(2$\times$1) surface. (b) Unpaired DB wire in $[\overline{1}10]$ direction on H:Si(100)-(2$\times$1) surface. (c) Unpaired DB wire in [110] direction on H:Si(100)-(2$\times$1) surface. (d) Part of the surface Brillouin zone used in the calculations. Values of ($k_1$, $k_2$) at different symmetry points ($\Gamma, J, K, J'$) in surface Brillouin zone are shown corresponding to $[\overline{1}10]$ and [110] directions respectively.}
\end{figure*}

\section{Atomic structure and assumptions} 

We use atomic structure of reconstructed Si(100)-(2$\times$1) surface as reported by Ramstad \textit{et al} \cite{Ramstad95} where the top four layers are relaxed due to surface reconstruction. Ramstad \textit{et al} \cite{Ramstad95} report structure for five layers of Si(100)-(2$\times$1)-AD surface and Si(100)-(2$\times$1)-SD surface. We add eleven bulk layers to make a sixteen layer structure and finally passivate the bottom layer with H in dihydride configuration as shown in Fig. 1(a) and (b) respectively. We then add H atoms to each of the Si dimer atoms of Si(100)-(2$\times$1)-SD structure, resulting in H:Si(100)-(2$\times$1) structure as shown in Fig. 1(c). Furthermore, we remove one of the H atoms from H:Si(100)-(2$\times$1) so that the resulting structure has an unpaired DB as shown in Fig. 1(d). Fig. 1(e) shows a H:Si(100)-(2$\times$1) surface with directions along and perpendicular to the dimer rows referred to as $[\overline{1}10]$ and [110] respectively. Apart from this, we add H atoms with H-Si bond length = 1.47$\AA$ and H-Si dimer angle = $111^\circ$. The bond length and the angle were obtained by structural optimization of a $Si_9H_{14}$ cluster using local spin density approximation with 6-311g* basis set as shown in Fig. 1(f) - calculations are performed using Gaussian\cite{G03}. The atomic co-ordinates for Si(100)-(2$\times$1)-AD unit cell (Fig. 1(a)) and H:Si(100)-(2$\times$1) unit cell (Fig. 1(c)) are given here\cite{AD,HSD}. These four unit cells, shown in Fig. 1(a-d), are used as building blocks to construct the bigger unit cell for doing isolated DB, DB wire and cluster calculations. In $[\overline{1}10]$ and [110] direction, the unit cell consists of varying layers of Si atoms depending upon the requirement. For example, to study an isolated unpaired DB, we need to include enough H:Si layers in $[\overline{1}10]$ and [110] direction so that the DBs in adjacent unit cells do not influence each other. We obtain this information from a previous study\cite{Raza06} that a DB affects neighboring atoms up to approximately 10$\AA$. Table I summarizes the dimensions of the unit cells and the number of atoms used to simulate different surfaces. Three representative examples of the unit cells are shown in Fig. 2. We show (4$\times$4), (4$\times$1) and (2$\times$4) unit cells in Fig. 2(a-c), which are used for calculating electronic band structure of an isolated unpaired DB and unpaired DB wire in $[\overline{1}10]$ and [110] directions respectively. Since, DBs interact electronically over a distance of approximately 10$\AA$, a (4$\times$4) unit cell makes certain that DBs in consecutive unit cells are electronically isolated. The calculated electronic structure is thus that of an isolated DB. Similarly, DB wires and DB clusters are also spaced sufficiently apart. Fig. 2(d) shows part of the surface Brillouin zone used in the calculation (with symmetry points ($\Gamma, J, K, J'$)\cite{SrivastavaBook}). 

For H:Si(100)-(2$\times$1), Cakmak \textit{et al} \cite{Cakmak03} report that the dimer distance is 2.35$\AA$ as compared to 2.23$\AA$ for Si(100)-(2$\times$1)-SD as obtained by Ramstad \textit{et al}\cite{Ramstad95} - suggesting an increase of approximately 5\%. However, surface states after passivation are well below $E_v$ and hence do not affect the band gap. It has also been reported elsewhere\cite{Cakmak03, Watanabe96} that an unpaired DB introduces a structural perturbation in H:Si(100)-(2$\times$1) structure, which is weak. We find that using this structure for Si dimer, the surface state is shifted by approximately 20meV. At room temperatue, this surface relaxation is anticipated to have small effect and hence in our calculations, we ignore it for simplicity. Furthermore, DB wires could go through Peierls distortion. It has however been reported \cite{Watanabe96} that the total energy gain due to such distortion is 14meV and is not anticipated to have a significant effect at room temperature. Moreover, Jahn-Teller distortion in DB clusters at low temperature has been discussed in Ref.\cite{Hitosugi99}. The effects due to these structural changes are also anticipated to be small at room temperature. Apart from this, the effect of dopant atoms and other defects (both surface and bulk) are ignored. All the atomic visulalizations in this paper are done using GaussView\cite{GW03}.

\begin{figure}
\vspace{2.5in}
%\centering\hskip -2.0in
\hskip -3.25in\includegraphics{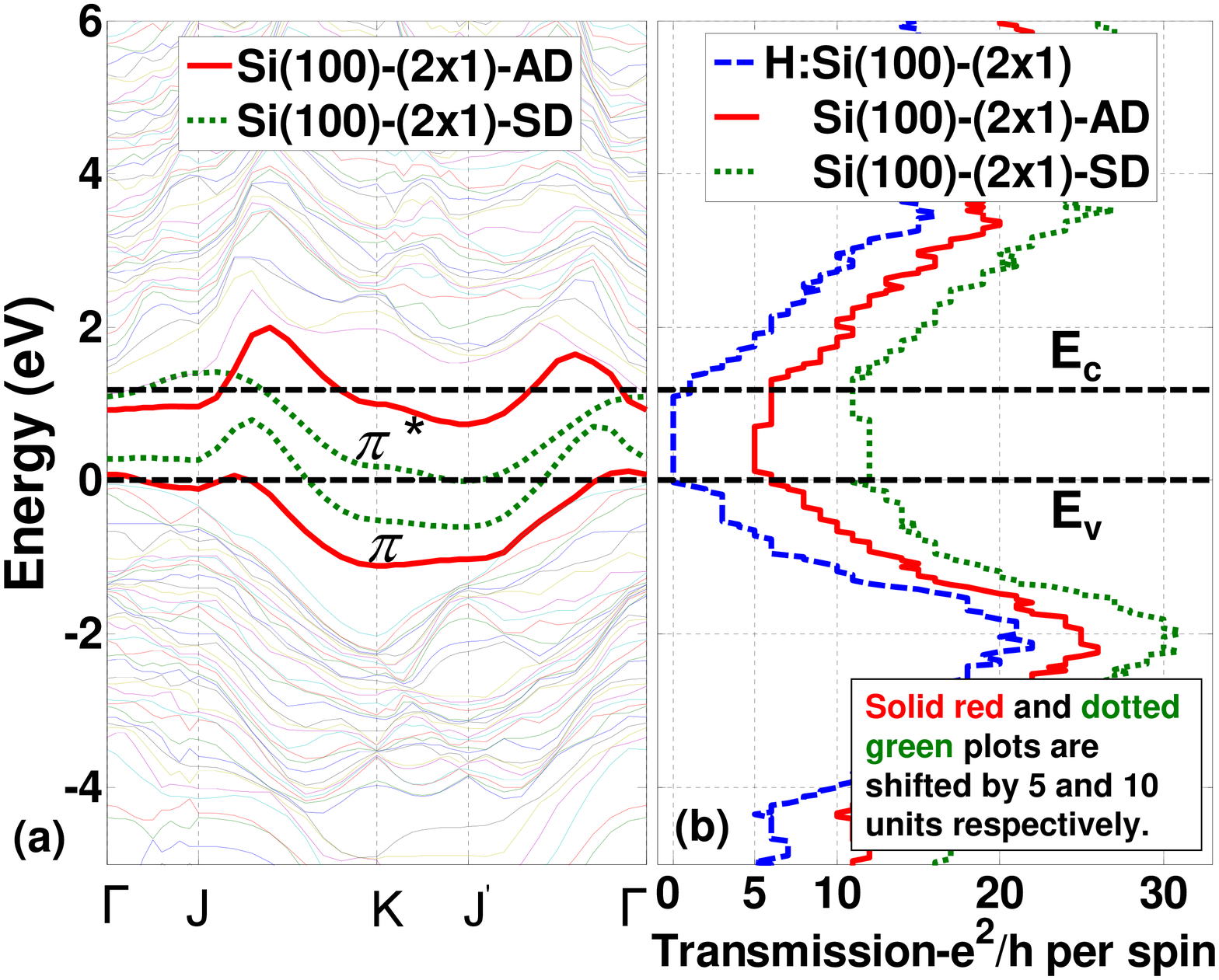}
\caption{(color online) Calculated electronic band structure and transmission plot for H: Si(100)-(2$\times$1),  Si(100)-(2$\times$1)-AD and Si(100)-(2$\times$1)-SD surface. (a) Band structure of H: Si(100)-(2$\times$1) surface shows clean band gap. All the states are shown for this surface. For  Si(100)-(2$\times$1)-AD and -SD surface, only $\pi$ and $\pi^*$ states are shown. For Si(100)-(2$\times$1)-AD surface, $\pi$ state is around 0.2eV below $E_v$ and $\pi^*$ state around 0.4eV below $E_c$ as expected \cite{Hamers93}. Similarly, $\pi$ and $\pi^*$ states of Si(100)-(2$\times$1)-SD surface spread through the band gap. (b) Transmission through different surfaces in units of ${e^2}$/$h$ per spin. Complementary view of part (a) reporting transmission through the above-mentioned three surfaces. H:Si(100)-(2$\times$1) surface has clean band gap. Si(100)-(2$\times$1)-AD and -SD surface has 0.6eV and no band gap respectively due to $\pi$ and $\pi^*$ states. The plots are shifted by 5 and 10 units along transmission axis for AD and SD surface respectively.}
\end{figure}

\section{Theoretical approach}

We model the Si surface by using a periodic system of unit cells repeating in $[\overline{1}10]$ and [110] directions. The lattice is defined by $p\overrightarrow{A_1}+q\overrightarrow{A_2}$, where $\overrightarrow{A_1}$ and $\overrightarrow{A_2}$ are lattice unit vectors in $[\overline{1}10]$ and [110] directions; p and q are indices representing the repeating unit cells in $\overrightarrow{A_1}$ and $\overrightarrow{A_2}$ directions respectively. The number of neighboring unit cells depend on the size of the unit cell. Hamiltonian (H) and overlap (S) matrices are computed in EHT scheme using parameters developed by Cerda \textit{et al} \cite{Cerda00}. With periodic boundary conditions in $[\overline{1}10]$ and [110] directions, Fourier transform technique is used to transform H and S from the real space to the reciprocal (k) space following the scheme as follows,

\begin{eqnarray}H(\overrightarrow{k})=\sum_{m=1}^N{H_{mn}e^{i\overrightarrow{k}.(\overrightarrow{d_m}-\overrightarrow{d_n})}}\end{eqnarray}
\begin{eqnarray}S(\overrightarrow{k})=\sum_{m=1}^N{S_{mn}e^{i\overrightarrow{k}.(\overrightarrow{d_m}-\overrightarrow{d_n})}}\end{eqnarray}

where $\overrightarrow{k}=\overrightarrow{k_1}+\overrightarrow{k_2}$, $\overrightarrow{k_1}$ and $\overrightarrow{k_2}$ are reciprocal lattice vectors of the surface Brillouin zone in $[\overline{1}10]$ and [110] direction; n represents the center unit cell; m represents the neighboring unit cells; N is the total number of unit cells; $\overrightarrow{d_m}-\overrightarrow{d_n}$ is the displacement between neighboring ($m^{th}$) and center ($n^{th}$) unit cell. Finally, we calculate energy Eigen values for different values of k along the surface Brillouin zone symmetry points ($\Gamma, J, K, J'$). 

\begin{figure}
\vspace{2.5in}
%\centering\hskip -2.0in
\hskip -3.25in\includegraphics{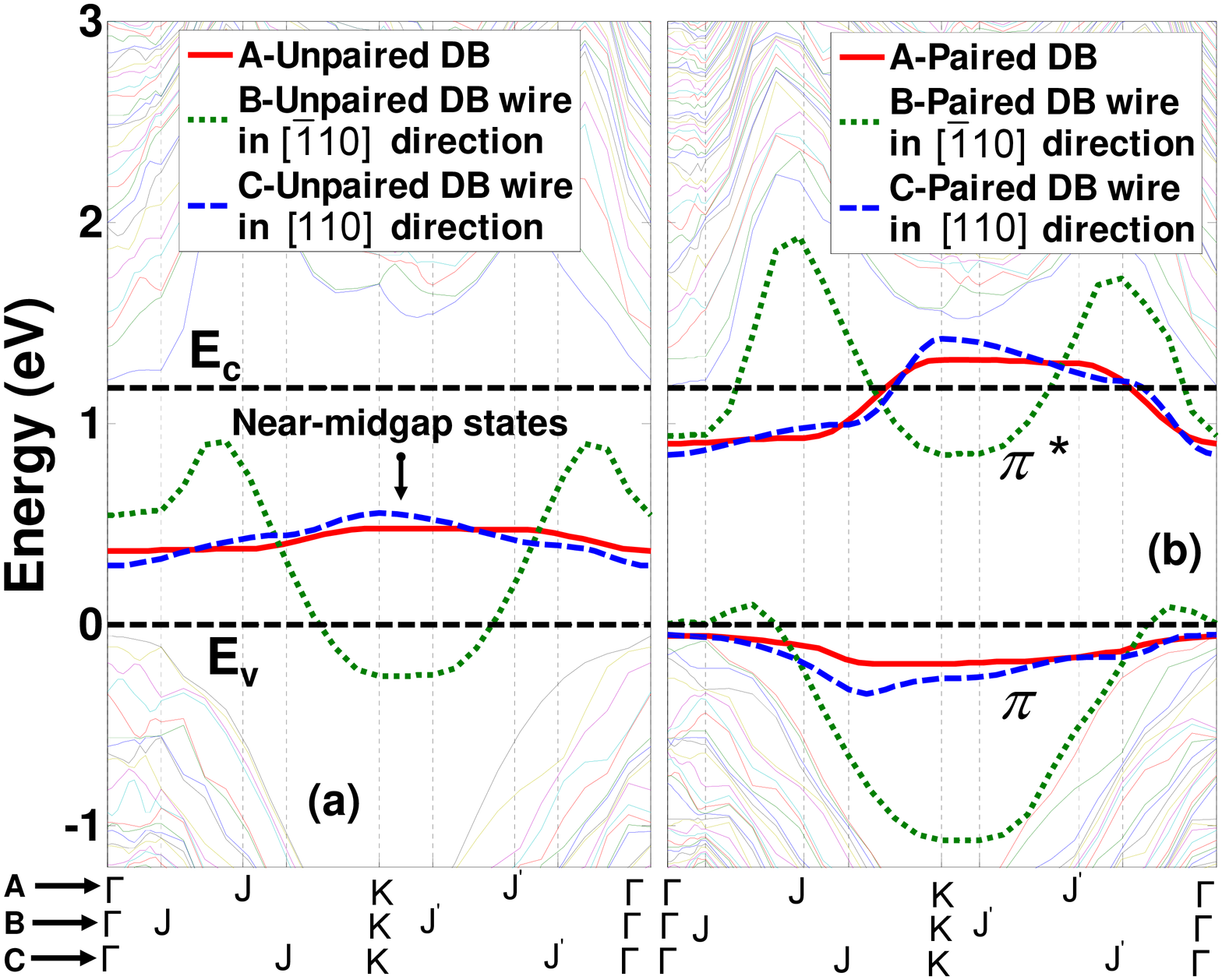}
\caption{(color online) Calculated electronic band structure of periodic unit cells containing isolated DBs and DB wires. (a) Band structure of periodic unit cells containing an unpaired DB and unpaired DB wires in $[\overline{1}10]$ and [110] direction. The unpaired DB introduces a near midgap state whose dispersion is small. The surface states due to an unpaired DB wire in $[\overline{1}10]$ direction have a large dispersion, due to DBs being 3.84$\AA$ apart. For the unpaired DB wire in [110] direction, DBs being farther apart (7.68$\AA$), thus the states are less dispersed. All the bulk and surface states for structure B are shown; whereas for structure A and C, states inside or close to band gap are shown. Refer to Fig. 2 for visualization of atomic structures of A, B and C. (b) Electronic band structure of periodic unit cells containing a paired DB and paired DB wires in $[\overline{1}10]$ and [110] direction. The paired DB introduces $\pi$ and $\pi^*$ states very similar to the ones introduced by Si(100)-(2$\times$1)-AD surface, however the dispersion is much smaller. The DBs in paired DB wire in $[\overline{1}10]$ are only 3.84$\AA$ apart and interact in a fashion similar to AD surface leading to considerable dispersion of the DB states. Similarly, for the paired DB wire in [110] direction, because the DBs are now farther apart, the surface states appear more like those of an isolated DB pair. All the bulk and surface states for structure B are shown; whereas for structure A and C, states inside or close to band gap are shown. The symmetry points ($\Gamma, J, K, J'$) along surface Brillouin zone for different configurations (A,B and C) are shown according to scale.}
\end{figure}

\begin{figure}
\vspace{2.5in}
%\centering\hskip -2.0in
\hskip -3.25in\includegraphics{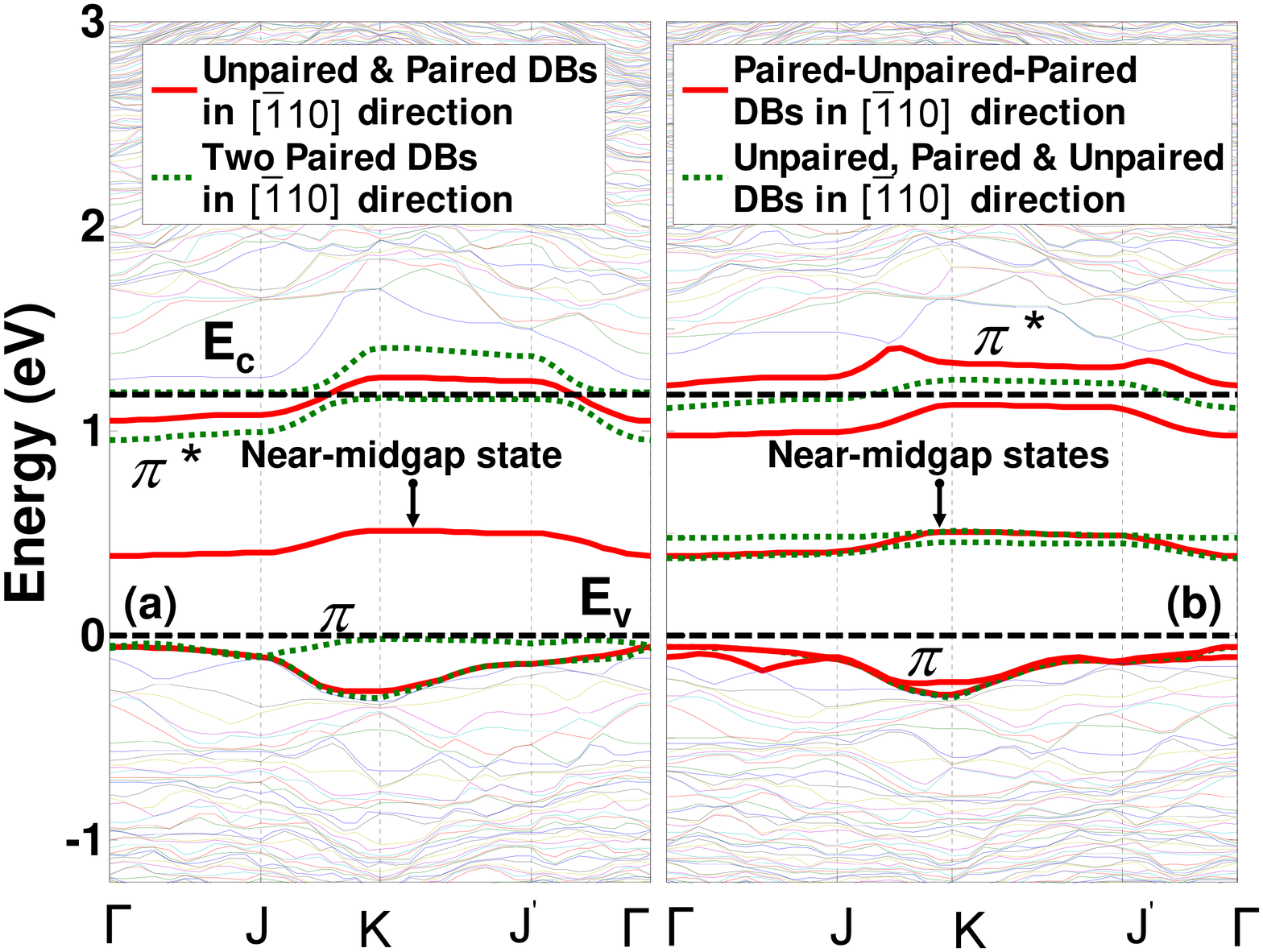}
\caption{(color online) Calculated electronic band structure of periodic unit cells containing different DB clusters. (a) Cluster of an unpaired and paired DBs in $[\overline{1}10]$ direction, and of two paired DBs in $[\overline{1}10]$ direction. All the bulk and surface states for the cluster of two paired DBs are shown; whereas for the cluster of unpaired and paired DBs, states inside or close to band gap are shown. (b) Cluster of paired-unpaired-paired DBs in $[\overline{1}10]$ direction and of unpaired-paired-unpaired DBs in $[\overline{1}10]$ direction. All the bulk and surface states for the cluster of paired-unpaired-paired DBs are shown; whereas for the cluster of unpaired-paired-unpaired DBs, states inside or close to band gap are shown.}
\end{figure}

\section{Results}

%In Fig. 3, we reproduce the results obtained by Liang \text{et al}\cite{Liang05} for H:Si(100)-(2$\times$1), Si(100)-(2$\times$1)-AD and Si(100)-(2$\times$1)-SD surfaces. By using a larger unit cell in [100] direction of sixteen Si layers, the quantum confinement effects are minimized.
Fig. 3(a) shows the electronic band diagram of H:Si(100)-(2$\times$1) surface showing the surface and bulk states resulting in a clean band gap of 1.18eV. For an Si(100)-(2$\times$1)-AD surface, it has been experimentally known that the $\pi$ state is 0.2eV below $E_v$ and $\pi^*$ state is 0.4eV below $E_c$\cite{Hamers93}. These $\pi$ and $\pi^*$ states for Si(100)-(2$\times$1)-AD surface are shown in Fig. 3(a), which are consistent with the experimental observation. The bandwidth of these states is 1.22eV and 1.26eV respectively. The band gap obtained in this case is 0.62eV. Similarly for Si(100)-(2$\times$1)-SD surface, the $\pi$ and $\pi^*$ states disperse through the whole band gap with bandwidths of 1.39eV and 1.42eV respectively. Fig. 3(b) gives a complementary view of Fig. 3(a) showing transmission through the above-mentioned three surfaces in the unit steps of $e^2/h$. 

Fig. 4 shows calculations for an isolated unpaired DB and unpaired DB wire in $[\overline{1}10]$ and [110] direction. An isolated unpaired DB introduces a near-midgap state\cite{Raza06} with a very small bandwidth of 0.11eV as shown in Fig. 4(a). We attribute this small dispersion to the energetic isolation of this state from the conduction and the valence band states. For the DB wire in $[\overline{1}10]$ direction, the bandwidth of the near-midgap state is 1.16eV, resulting in a very small band gap as shown in Fig. 4(a). We interpret this large dispersion due to the DBs being in the close proximity of 3.84$\AA$ and hence interacting strongly. For the DB wire in [110] direction, the distance between DBs is 7.68$\AA$. Since DBs are farther apart in this case, the bandwidth is 0.26eV. This bandwidth is much smaller than that of the wire in $[\overline{1}10]$ direction and larger than an isolated DB as shown in Fig. 4(a). 

Fig. 4(b) presents calculation for an isolated paired DB and paired DB wire in $[\overline{1}10]$ and [110] direction. The electronic structure of a paired DB is significantly different from an isolated unpaired DB. It introduces a bonding ($\pi$) and anti-bonding ($\pi^*$) state below $E_v$ and up to 0.28eV below $E_c$ respectively. Qualitatively, this behavior is similar to unpassivated reconstructed Si surface. However, dispersion of these states is much smaller. The bandwidth of this $\pi$ and $\pi^*$ state is 0.14eV and 0.42eV respectively. The electronic band structure for a paired DB wire in $[\overline{1}10]$ direction shows the similar $\pi$ and $\pi^*$ states with bandwidths of 1.17eV and 1.08eV respectively. This large dispersion is due to strong coupling between the DBs in this configuration, being 3.84$\AA$ apart. As expected, the dispersion of $\pi$ and $\pi^*$ states for DB pair wire in [110] direction is less than that of $[\overline{1}10]$ direction and more than that of an isolated DB pair. The bandwidth is 0.29eV and 0.57eV for $\pi$ and $\pi^*$ state respectively. For an isolated paired DB and paired DB wire along [110] direction, only the state inside or close to band gap are shown, whereas, for the paired DB wire along $[\overline{1}10]$ direction all the bulk and surface states are shown. Since the dimensions of the unit cell for configurations A, B and C are different, the symmetry points ($\Gamma, J, K, J'$) along the surface Brillouin zone are labeled accordingly. Qualitatively, the results reported are similar to those of Watanabe \textit{et al}\cite{Watanabe96} for unpaired DB wire in $[\overline{1}10]$ direction and unpaired and paired DB wire in $[\overline{1}10]$ and [110] directions. However, there are some differences in the shape of dispersion for the paired DB wire in these directions. 

Fig. 5 presents electronic band structure calculation for DB clusters. The qualitative behavior can be attributed to a combined behavior of individual unpaired and paired DBs, which are 3.84$\AA$ apart. Fig. 5(a) shows states due to a cluster of unpaired and paired DBs, 3.84$\AA$ apart. The resulting states consist of one near-midgap state with bandwidth of 0.12eV and two $\pi$ and $\pi^*$ state with bandwidth of 0.26eV and 0.22eV respectively. Qualitatively, we interpret these states as a combination of states due to an unpaired and paired DB respectively. However, the dispersion is larger, which we attribute to the mutual couplings of these unpaired and paired DBs. Similarly, states due to a cluster of two paired DBs, 3.84$\AA$ apart, result in two $\pi$ and two $\pi^*$ states, as shown in Fig. 5(a). These states are similar to the ones due to isolated paired DB. However, one of these $\pi$ states (plotted as green dotted line) has very small bandwidth of 83meV that needs to be explored further. The bandwidth of other $\pi$ state is 0.25eV and that of two $\pi^*$ states is 0.2eV and 0.22eV. Fig. 5(b) extends the cluster size by incorporating two clusters of (i) paired-unpaired-paired DBs, each 3.84$\AA$ away, and (ii) unpaired-paired-unpaired DBs, each 3.84$\AA$ away. Qualitatively, the behavior can again be interpreted as combination of surface states due to individual paired and unpaired DBs. However, we again find a $\pi$ state of relatively small bandwidth of 76meV that needs further exploration. 

%\begin{figure}
%\vspace{2.2in}
%\centering\hskip -2.0in
%\hskip -2.75in\special{psfile=Fig6.eps hscale=28.0 vscale=28.0}
%\caption{(color online) Calculated band structure of periodic unit cells containing isolated DB clusters. (a) 1DB + 2DB in $[\overline{1}10]$ direction. (b)   \textcolor{red}{ALRIGHT !!!}}
%\end{figure}

\section{Conclusions}

We have reported the electronic band structure calculations showing differences between states induced in Si band gap due to isolated DB, DB wires and clusters. We extend on previous studies and report that the details of these states inside band gap vary considerably from isolated DBs to DB wires and clusters. An unpaired DB behaves completely differently from a paired DB. Similarly, isolated DBs are different from DB wires and clusters. Furthermore, we report that an isolated unpaired DB and paired DB have a very small dispersion in their induced states. However, wires introduce a larger dispersion that can be attributed to the additional coupling between the DBs. Similarly, DB clusters broaden the induced states depending upon the coupling between the individual unpaired and paired DBs. We have ignored contribution of dopant atoms, dephasing processes, further surface relaxation effects and charge transfer effects in our calculations. 

We thank S. Datta, G.-C. Liang, T. Raza, D. Kienle and K. H. Bevan for useful discussions. This work was supported by the NASA Institute for Nanoelectronics and Computing (INAC). Computational facilities were provided by the NSF Network for Computational Nanotechnology and nanoHUB.org\cite{Nanohub}.
%S. Datta, G.-C. Liang, T. Raza, D. Kienle and Kirk H. Bevan


\begin{thebibliography}{100}
\bibitem{AD} Atomic co-ordinates for Si(100)-(2$\times$1)-AD unit cell - (Si, 1.16200, 0.00000, -0.92100); (Si, 3.30559, 0.00000, -0.21300); (Si, 0.06600, 1.91979, -1.49850); (Si, 3.74059, 1.91979, -1.46950); (Si, 1.95079, 1.91979, -2.95500); (Si, 5.73438, 1.91979, -2.71800); (Si, 1.90679, 0.00000, -4.22750); (Si, 5.75438, 0.00000, -4.07050). The remaining twenty four Si and four H atoms have same co-ordinates as those of the last twenty four Si and four H atoms of the H:Si(100)-(2$\times$1) unit cell\cite{HSD}.
\bibitem{Nanohub} nanoHUB.org is a web site dedicated to advancing nanotechnology through theory, modeling and simulation for research and education.
\end{thebibliography}
\end{document}